\documentclass{article}
\usepackage{amssymb}

\usepackage{graphicx}
\usepackage{amsmath}

\input{tcilatex}

\begin{document}

\title{$S$-wave $\pi -\pi $ scattering lengths\\
}
\author{N. F. Nasrallah \\
Faculty of Science, Lebanese University\\
Tripoli, Lebanon}
\date{}
\maketitle

\begin{abstract}
An old calculation of the $s$-wave $\pi -\pi $ scattering lengths is updated
and supplemented with experimental data on the $\pi -\pi $ $s$-wave phase
shift. The results 
\begin{equation*}
ma_{0}=.215\text{ \ \ and \ \ }ma_{2}=-.039
\end{equation*}
are in excellent agreement with those obtained from a recent analysis of the
results published by the Brookhaven \textrm{E865} collaboration.\bigskip
\bigskip 
\end{abstract}

The subject of the $s$-wave $\pi -\pi $ scattering lengths has witnessed two
main peaks of both experimental and theoretical activity since they were
first calculated by Weinberg \cite{sweinberg} in 1966. On the experimental
side the Geneva-Saclay experiment \cite{lrosselet} carried in the mid 1970's
gave for the isoscalar scattering length $ma_{0}=.26\pm .05.$ Recently a new
measurement of $K_{e4}$ decay and of the $\pi -\pi $ phase-shift difference $%
\delta _{0}^{0}-\delta _{1}^{1}$ has been published by the Brookhaven 
\textrm{E865} collaboration \cite{pislak}\ with statistics improved by more
than a factor of $10$. Analysis and interpretation of the new data has
recently been carried out by Descotes et al. \cite{descotes} who used
solutions of the Roy equations \cite{roy}\ obtained by Ananthanarayan et al. 
\cite{anan}. Their results for the scattering lengths are 
\begin{equation}
ma_{0}=.228\pm .012\ \ \ \ \ \ \ ,\ \ \ \ \ \ ma_{2}=-.0382\pm .0038
\label{1}
\end{equation}

On the theoretical side considerable activity was devoted to the subject in
the late 1960's \cite{schwinger}. Subsequently the effective low energy
theory of \textrm{QCD}, Chiral Perturbation Theory (\textrm{ChPT}) was
applied to the problem. In \textrm{ChPT} the scattering amplitude is
expanded in powers of the momenta and of the quark masses. \textrm{ChPT} is
a non renormalizable theory and subtraction constants (low energy couplings)
have to be introduced at each order of the calculation. The elaborate
evaluation of the perturbation series to two loops was completed only
recently \cite{bijnens}, \cite{clangelo}. Four low energy couplings $l_{1},$ 
$l_{2},$ $l_{3},$ $l_{4}$ enter in the calculation the values of $l_{1}$ and 
$l_{2}$ were obtained in ref. \cite{clangelo}\ by solving the Roy equations 
\cite{roy}. $l_{3}$ and $l_{4}$ on the other hand are obtained only
indirectly. $l_{4}$ is expressible in terms of the scalar radius of the
pion. 
\begin{equation}
r_{s}^{2}=.61\pm .04\,\mathrm{fm}^{2}  \label{2}
\end{equation}
obtained from an analysis of the s-wave isoscalar phase shifts \cite{donohue}%
. $l_{3}$ is related to the variation of the pion mass $m$ from its chiral
limit. The result obtained in ref. \cite{clangelo}\ for the scattering
lengths is 
\begin{equation}
ma_{0}=.220\pm .005\ \ \ \ \ \ \ ,\ \ \ \ \ \ \ ma_{2}=-.0444\pm .0010
\end{equation}

It is the purpose of this note to point out that the method of collinear
dispersion relations \cite{fubini}\ as used in \cite{nasrallah}\ is
particularly well adapted to the problem of the $\pi -\pi $ $s$-wave
scattering lengths. It provides a simple alternative to \textrm{ChPT} and
yields 
\begin{equation}
ma_{0}=.215\ \ \ \ \ \ \ ,\ \ \ \ \ \ ma_{2}=-.039
\end{equation}
when supplemented with experimental data on the s-wave isoscalar
phase-shifts which are used to estimate the variation of the pion isoscalar
form factor between momentum transfers $s=0$ and $s=4m^{2}.$

We start from the expression 
\begin{equation}
T_{\mu \nu }=\frac{i}{f_{\pi }^{2}m^{4}}\int
d^{4}ye^{ip_{a}y}(m^{2}-p_{a}^{2})(m^{2}-p_{b}^{2})\langle \pi ^{c}\left|
TA_{\mu }^{a}(y)A_{\nu }^{b}(0)\right| \pi ^{d}\rangle
\end{equation}
where $f_{\pi }=.0924\mathrm{GeV}$ is the pion decay constant and $A_{\mu
}^{i}=1/2\overline{q}\lambda ^{i}\gamma _{\mu }q$ denote the axial-vector
currents.

The method of collinear dispersion relations \cite{fubini}\ consists of
writing a dispersion relation in the collinear variable $x$ in the rest
frame of the target pion where 
\begin{equation}
p_{c}=p_{d}=p\ \ \ \ \ \ \ \ \ ,\ \ \ \ \ \ \ \ p_{a}=p_{b}=xp=q
\end{equation}
Current Algebra and the generalised Ward-Takahashi identity give 
\begin{equation}
x^{2}p_{\mu }p_{\nu }T_{\mu \nu }=U(x)-\frac{2m^{2}}{f_{\pi }^{2}}%
.x.(1-x^{2})^{2}.(\delta _{ac}\delta _{bd}-\delta _{ad}\delta _{bd})+\delta
_{ab}\delta _{cd}S  \label{7}
\end{equation}
with 
\begin{equation}
U(x)=\frac{i}{f_{\pi }^{2}m^{4}}\int d^{4}ye^{iqy}(m^{2}-q^{2})^{2}\langle
\pi ^{c}\left| TD_{a}(y)D_{b}(0)\right| \pi ^{d}\rangle
\end{equation}
\begin{equation}
S=\frac{i}{f_{\pi }^{2}}\langle \pi \left| \left[ Q,D(y)\right]
_{e.t}\right| \pi \rangle
\end{equation}
where 
\begin{equation}
D=\partial _{\mu }A_{\mu },Q=\int d^{3}yA(\overset{\rightarrow }{y},0)
\end{equation}
Eq. (\ref{7}) yields 
\begin{equation}
\underset{x\rightarrow o}{\lim }U(x)=-\delta _{ab}\delta _{cd}S
\end{equation}
\begin{equation}
\underset{x\rightarrow 0}{\lim }\frac{dU(x)}{dx}=-\frac{2m^{2}}{f_{\pi }^{2}}%
(\delta _{ad}\delta _{bc}-\delta _{ac}\delta _{bd})
\end{equation}
it is convenient to perform the isospin decomposition of $U(x)$ as 
\begin{equation}
U(x)=A(x)\delta _{ab}\delta _{cd}+B(x)(\delta _{ad}\delta _{bc}-\delta
_{ac}\delta _{bd})+x.C(x).(\delta _{ad}\delta _{bc}+\delta _{ac}\delta _{bd})
\end{equation}
with $A$, $B$ and $C$ even under crossing: $A(x)=A(-x)$, etc... It also
follows from the definition of $U$ that 
\begin{equation}
U(x=1)=T_{th}
\end{equation}
$T_{th}$ denoting the transition matrix element of the process $\pi ^{b}+\pi
^{d}\rightarrow \pi ^{a}+\pi ^{c}$ at threshold.

Bose symmetry imposes 
\begin{equation}
A(x=1)=B(x=1)+C(x=1)  \label{14}
\end{equation}
and in the soft pion limit 
\begin{eqnarray}
A(x &=&0)=-S  \notag \\
B(x &=&0)=0 \\
C(x &=&0)=-\frac{2m^{2}}{f_{\pi }^{2}}  \notag
\end{eqnarray}

In the complex $x$-plane, $A$, $B$ and $C$ are analytic functions of $x$
with cuts on the real axis extending from $-\infty $ to $-1$ and from$1$ to $%
\infty .$ By Cauchy's theorem then 
\begin{eqnarray}
A(x &=&1)=-S+\frac{1}{2\pi i}\int_{c}\frac{dx}{x(x^{2}-1)}A(x)  \notag \\
B(x &=&1)=\frac{1}{2\pi i}\int_{c}\frac{dx}{x(x^{2}-1)}B(x)  \label{16} \\
C(x &=&1)=\frac{-2m^{2}}{f_{\pi }^{2}}\int_{c}\frac{dx}{x(x^{2}-1)}C(x) 
\notag
\end{eqnarray}
where c is the contour consisting of straight lines parallel to the real
axis immediatly above and below the cuts and a circle of large radius $R.$
The integrals over the circles in the expression above are determined by the
asymptotic behaviour of the functions $A$, $B$ and $C.$ On the upper and
lower arcs $x^{2}$ is large and negative so that the operator product
expansion for the time ordered product can be used. This constitutes a good
approximation except in the vicinity of the real axis. We have 
\begin{equation}
\int dye^{iqy}TD_{a}(y)D_{b}(0)\underset{x^{2}\rightarrow -\infty }{%
\longrightarrow }C_{1}+\frac{O_{1}}{x^{2}}+...,O_{1}\varpropto \overline{q}q
\end{equation}
The contribution of the unit operator (perturbative) vanishes because only
connected parts of the amplitude enter. the integrals over the circles in
Eq. (\ref{16}) amount to $\langle \pi ^{c}\left| O_{1}\right| \pi
^{d}\rangle \thicksim \frac{m_{q}^{2}}{m_{\pi }^{2}}S$ and are hence
negligible. To a good approximation then we can rewrite Eq. (\ref{16})
making use of crossing symmetry 
\begin{mathletters}
\begin{eqnarray}
A(x &=&1)=-S+\frac{2}{\pi }\int_{1}^{\infty }\frac{dx}{x(x^{2}-1)}Abs\,A(x)
\label{19a} \\
B(x &=&1)=\frac{2}{\pi }\int_{1}^{\infty }\frac{dx}{x(x^{2}-1)}Abs\,B(x)
\label{19b} \\
C(x &=&1)=\frac{-2m^{2}}{f_{\pi }^{2}}+\frac{2}{\pi }\int_{1}^{\infty }\frac{%
dx}{x(x^{2}-1)}Abs\,C(x)  \label{19c}
\end{eqnarray}
The isospin amplitudes at threshold are linear combinations of the
expresions above 
\end{mathletters}
\begin{equation}
T_{0,2}^{th}=U_{0,2}(x=1)
\end{equation}
where 
\begin{eqnarray}
U_{0}(x) &=&5B(x)-c(x)  \notag \\
U_{2}(x) &=&2(B(x)-C(x))  \label{21}
\end{eqnarray}

In addition the threshold behaviour of the amplitudes is 
\begin{eqnarray}
T_{0,2}^{th} &=&32\pi ma_{0,2}  \notag \\
\underset{x\rightarrow 1}{\lim }Abs\,U_{0,2}(x) &=&\underset{x\rightarrow 1}{%
\lim }Abs\,T_{0,2}(x)=32\pi m^{2}a_{0,2}^{2}.\sqrt{(x^{2}-1)}
\end{eqnarray}
The second of the expressions above contributes an integrable threshold
singularity to the integrals appearing in Eq.(19). Eqs.(\ref{19b}) and (\ref
{19c}) together with Eq. (\ref{21}) yield then 
\begin{eqnarray}
32\pi ma_{0} &=&\frac{2m^{2}}{f_{\pi }^{2}}+32\pi m^{2}a_{0}^{2}+\frac{2}{%
\pi }\int_{1}^{\infty }\frac{dx}{x(x^{2}-1)}Abs\,(U_{0}(x)-U_{0}(1))  \notag
\\
&&  \label{23} \\
32\pi ma_{2} &=&-\frac{4m^{2}}{f_{\pi }^{2}}+32\pi m^{2}a_{2}^{2}+\frac{2}{%
\pi }\int_{1}^{\infty }\frac{dx}{x(x^{2}-1)}Abs\,(U_{2}(x)-U_{2}(1))  \notag
\end{eqnarray}
Because of the constraint Eq.(\ref{14}) we also have from Eqs.(\ref{19a})
and (\ref{21}) 
\begin{equation}
32\pi ma_{2}=-2S+32\pi m^{2}a_{2}^{2}+\frac{2}{\pi }\int_{1}^{\infty }\frac{%
dx}{x(x^{2}-1)}Abs\,(A(x)-A(1))  \label{24}
\end{equation}
Using the reduction technique, $Abs\,U$ can be decomposed into three parts 
\begin{equation}
Abs\,U_{1}=\frac{(2\pi )^{4}}{2}\underset{n}{\sum }\langle \pi ^{c}\left|
j_{a}\right| n\rangle \langle n\left| j_{b}\right| \pi ^{d}\rangle \delta
(p+q-p_{n})-\left( a\longleftrightarrow b,q\longleftrightarrow -q\right) 
\notag
\end{equation}
\begin{equation*}
Abs\,U_{2}=\frac{(2\pi )^{4}}{2}\underset{m}{\sum (}\langle 0\left|
j_{a}\right| m\rangle \langle m,\pi ^{c}\left| j_{b}\right| \pi ^{d}\rangle
\end{equation*}
\begin{equation}
+\langle \pi ^{c}\left| j_{a}\right| \pi ^{d},m\rangle \langle m\left|
j_{b}\right| 0\rangle )\delta (q-p_{m})-(a\longleftrightarrow
b,q\longleftrightarrow -q)  \label{25}
\end{equation}
\begin{equation*}
Abs\,U_{3}=\frac{(2\pi )^{4}}{2}\underset{l}{\sum }\langle 0\left|
j_{a}\right| l,\pi ^{d}\rangle \langle l,\pi ^{c}\left| j_{b}\right|
0\rangle \delta (q-p-p_{l})-(a\longleftrightarrow b,q\longleftrightarrow -q)
\end{equation*}
Where $j=(\Box +m^{2})D$ and where only connected parts of the matrix
elements enter.

$Abs\,U_{1}$ contains the usual singularities in the $s,\overset{-}{s}$
channels. $Abs\,U_{2}$ contains the mass singularities associated with the
vertices $\langle 0\left| j\right| m\rangle $ and $Abs\,U_{3}$ coresponds to
the so-called $Z$ graphs.

The contribution of the single pion state to $AbsU_{2}$ vanishes
identically,the contribution of the $0^{-}$ continuum ,the same that
provides the corrections to the Goldberger-Treiman relation ,is strongly
damped and is not expected to amount to more than a few percent of the total
and shall be neglected.

The isoscalar channel ($\sigma (600)$, $f_{0}(980)$, $f_{0}(1370)$,$\cdots $%
) is expected to practicaly saturate the contribution of the $0^{+}$ $n$ and 
$l$ intermediate states.It is readily seen that these states contribute only
to $B(x)$ an amount which we denote by $b.$ We thus have from Eqs. (\ref{23}%
) and (\ref{24}) 
\begin{eqnarray}
32\pi ma_{0} &=&\frac{2m^{2}}{f_{\pi }^{2}}+32\pi m^{2}a_{0}^{2}+5b  \notag
\\
32\pi ma_{2} &=&-\frac{4m^{2}}{f_{\pi }^{2}}+32\pi m^{2}a_{2}^{2}+2b \\
32\pi ma_{2} &=&-2S+32\pi m^{2}a_{2}^{2}  \notag
\end{eqnarray}
If the contribution $b$ of the continuum were negligible we would deduce
from the last two equations above that 
\begin{equation}
S\backsimeq \frac{2m^{2}}{f_{\pi }^{2}}  \label{27}
\end{equation}
and corresponding values for the scattering lengths which come out close to
the ones obtained by Schwinger and by Balachandran et. al \cite{schwinger}.

Recall however 
\begin{equation}
S=\langle \pi \left| \sigma \right| \pi \rangle ,\sigma =\frac{i}{f_{\pi
}^{2}}\left[ Q,D\right] _{e.t}
\end{equation}
and consider 
\begin{eqnarray}
M(x) &=&\frac{i}{f_{\pi }m^{2}}\int d^{4}ye^{iqy}(m^{2}-q^{2})\langle
0\left| D(y)\sigma (0)\right| \pi (p)\rangle \\
M(x &=&1)=S
\end{eqnarray}
where the collinear parametrization $q=px$ has been used once again. The
soft pion limit now reads 
\begin{equation}
M(x=0)=-\frac{i}{f_{\pi }}\langle 0\left| \left[ Q,\sigma \right]
_{e.t}\right| \pi \rangle =\frac{m^{2}}{f_{\pi }^{2}}
\end{equation}

$M(x)$ is an analytic function in the cut complex $x$-plane in particular in
the interval $-1\leq x\leq 3$. We thus have to $O(m^{2})$%
\begin{equation}
M(x=0)=\frac{1}{2}.(M(x=1)+M(x=-1))
\end{equation}
$M(x=-1)$ represents the amplitude $\langle 0\left| \sigma \right| \pi ,\pi
\rangle _{th.}$ We have then 
\begin{equation}
S=\frac{m^{2}}{f_{\pi }^{2}}.(1-\delta )
\end{equation}
with 
\begin{equation}
\delta =\frac{M(x=-1)-M(x=1)}{M(x=-1)+M(x=1)}=O(m^{2})
\end{equation}
so that 
\begin{equation}
b=\frac{m^{2}}{f_{\pi }^{2}}.(1+\delta )
\end{equation}
It is clear from the above that Eq.(\ref{27}) cannot hold and that 
\begin{eqnarray}
32.\pi .ma_{0} &=&\frac{7m^{2}}{f_{\pi }^{2}}+32.\pi .m^{2}a_{0}^{2}+\frac{%
5m^{2}}{f_{\pi }^{2}}.\delta  \notag \\
32.\pi .ma_{2} &=&-\frac{2m^{2}}{f_{\pi }^{2}}+32.\pi .m^{2}a_{2}^{2}+\frac{%
2m^{2}}{f_{\pi }^{2}}.\delta
\end{eqnarray}

A remark is here in order: to lowest order in $m^{2}$ we recover the results
of Weinberg \cite{sweinberg} 
\begin{equation}
ma_{0}=\frac{7m^{2}}{32.\pi .f_{\pi }^{2}},ma_{2}=-\frac{m^{2}}{16.\pi
.f_{\pi }^{2}}
\end{equation}

The corrections to the chiral limit thus arise from two sources : a major
one proportional to $m^{2}a_{0,2}^{2}$ coming from the threshold singularity
and a minor one due to the variation $\delta $ of the pion scalar form
factor between momentum transfers $s=0$ and $s=4m^{2}.$ The results turn out
to be quite insensitive to the exact value of $\delta $ a reliable estimate
of which can be inferred assuming elastic unitarity for the pion scalar form
factor and using the well known solutions of the Muskhelishvili-Omnes
equations \cite{muskhelishvili}\ which give 
\begin{equation}
\frac{M(x=-1)}{M(x=1)}=\exp (\frac{4m^{2}}{\pi }\int_{4m^{2}}^{\infty }\frac{%
\delta _{0}(s).ds}{s(s-4m^{2})})-1  \label{37}
\end{equation}

$\delta _{0}$ is the isoscalar s-wave $\pi -\pi $ phase shift experimental
measurements of which are available up to $s_{1}\backsimeq .8\mathrm{GeV}$ 
\cite{lrosselet}, \cite{protopopescu}. Taking $\delta _{0}(s\geq
s_{1})=\delta _{0}(s_{1})$ should yield a good approximation for the the
rapidly convergent\ integral appearing in Eq. (\ref{37}). This gives 
\begin{equation}
\delta =.10
\end{equation}
$\delta \backsimeq \frac{4m^{2}}{6}.r_{s}^{2}$ with $r_{s}^{2}$ given by Eq.
(\ref{2}) results in practically the same value.

We finally obtain for the scattering lengths 
\begin{equation}
ma_{0}=.215,ma_{2}=-.039  \label{39}
\end{equation}

The error introduced in the numbers above by neglecting the contribution of
the $0^{-}$ continuum in Eq. (\ref{25}) should not amount to more than a few
percent. That the error due to the neglect of the contribution of the
non-threshold intermediate states is small is supported by the fact that no
structure is reported in this channel \cite{pdg}.This is confirmed by the
excellent agreement between Eq. (\ref{39}) and Eq. (\ref{1}).

We conclude from the above that the method of collinear dispersion
relations.provides a simple and reliable alternative to ChPT in thes-wave $%
\pi -\pi $ scattring sector and that new low energy $\pi -\pi $ scattering
data are more than welcome \cite{adeva}.\pagebreak

\end{document}